# The Kinetic Basis of Self-Organized Pattern Formation

Yuri Shalygo[1]

[1]Gamma Ltd, Vyborg, Russia
yuri.shalygo@gmail.com

## Abstract

In his seminal paper on morphogenesis (1952), Alan Turing demonstrated that different spatio-temporal patterns can arise due to instability of the homogeneous state in reaction-diffusion systems, but at least two species are necessary to produce even the simplest stationary patterns. This paper is aimed to propose a novel model of the analog (continuous state) kinetic automaton and to show that stationary and dynamic patterns can arise in one-component networks of kinetic automata. Possible applicability of kinetic networks to modeling of real-world phenomena is also discussed.

## Introduction

Pattern formation has been a well-investigated research field since the pioneering work of Alan Turing on morphogenesis (Turing, 1952). He showed that pattern formation can be accomplished by the interaction of two substances that spread with different rates. For the interactions of this type, Turing introduced the term reaction-diffusion systems, which is now generally in use. He demonstrated that a homogeneous steady state is unstable in certain such systems, and any small local deviation from this steady state is sufficient to trigger the onset of pattern formation. This phenomenon, termed diffusion-driven instability, has been found to be applicable in many biological and chemical systems and was analyzed mathematically using a variety of techniques (Murray, 1993).

The classical Turing instability leads to the establishment of stationary spatial patterns. However, the oscillatory analogue of this instability is possible and it was also envisaged by Alan Turing. This oscillatory instability produces traveling or standing waves and therefore it is often called the wave instability (Walgraef, 1997). Turing suggested that at least three species are needed for the oscillatory instability. The oscillatory instability is much rarer and has been found only in some chemical systems (Zhabotinsky et al., 1995) and biological phenomena (Meinhardt, 2004).

Owing to Turing's indisputable authority, it is often considered as a well-established fact that patterns can arise only in multicomponent dynamical systems. Actually, it holds mainly for reactive systems with passive transport, or, more precisely, chemical continuum media with gradient driven diffusion, described by Fick's Law and explained by Einstein in 1905 under the random walk assumption. However, an increasing number of natural phenomena, which are called sub-diffusion, super-diffusion or anomalous diffusion in general (Klafter and Sokolov, 2005), do not fit this relatively simple description of diffusion. From the signaling of biological cells to the foraging behaviour of animals, the overall motion of objects is better described by steps that are not independent and can take vastly different time to perform. Some systems can be modeled as networks of decision-making agents reacting to external events or signals and can be regarded as reflexive systems with active transport, where the direction of motion depends not only on gradients but on many other factors, which may lead to "The rich get richer" phenomenon.

This paper is aimed to propose a novel numerical algorithm, called *Conservative Rank Transform* and an analog model of an abstract autonomous agent, called a *kinetic automaton* or a *kinon* for short; then to show that different dynamical patterns found in Turing multi-component systems can arise even in one-component *kinetic networks*, defined as *reflexive dynamical systems with active transport*.

## Background

The proposed model stems from the cellular automata (CA) framework, conceived in the early 1950's by J. Von Neumann (Neumann, 1960) and Stanislaw Ulam (Ulam, 1952) and became popular among researchers largely due to John Conway's Game of Life (Gardner, 1970). The popularity of CA can be attributed to the enormous potential they hold in modeling complex systems and their simplicity, which is determined by the regularity of an underlying lattice and a fixed number of cell states. Frisch, Hasslacher and Pomeau (Frisch et al., 1986) and Wolfram (Wolfram, 1986) independently discovered that a simple cellular automaton on a 2D triangular lattice can simulate the Navier-Stokes equations and proposed an FHP model or Lattice Gas Automata (LGA), as these models are usually termed. Closely related to CA, Random Boolean Networks (RBN) were introduced by Stuart Kauffman as a model of genetic regulatory networks (Kauffman, 1969). It has been shown that Boolean idealization may capture the dynamics of genetic regulatory systems (Kauffman, 1993), but in general, Boolean approximation is inappropriate for modeling flow processes, e.g., movement of money through an economy, electricity over a grid, concentrations of metabolites in cell tissues, etc.

To address this issue, Coupled Map Lattices (CML) were proposed by Kunihiko Kaneko as a paradigm for the study of spatio-temporal complexity such as turbulence, convection, flows, population dynamics, etc. (Kaneko, 1985). CML can be viewed as a generalization of CA in terms of continuous state space and arbitrary network topology, but despite promising universality, CMLs have not become widespread.

On the contrary, Lattice Boltzmann Model (LBM), originally evolved from LGA and based on a minimal kinetic Boltzmann equation (Wolf-Gladrow, 2000), is attracting growing popularity. In LBM, representative particles ('parcels of fluid') evolve on a regular grid in accordance with simple streaming and collision rules designed to preserve fluid dynamics.

In the recent years, a lot of research has been directed to the continualization of the state space of different models. It was shown by Ulam that many deterministic problems in mathematics and physics can be converted into equivalent random processes and described by probabilities, which are real numbers (Ulam, 1952). The superiority of continuous values is confirmed by the mathematical theory of computable numbers and computable functions. It has been proved that simple analog (continuous state) computers can compute numbers and functions which are not computable by digital computers (Pour-El and Richards, 1989).

Despite the shown diversity, almost all existing modeling paradigms are derived from the cellular automaton model and inherit its restriction to lattice uniformity and a discrete state space. In some cases it leads to excessive oversimplification. According to Albert Einstein, *"Everything should be made as simple as possible, but no simpler."* The cellular automaton paradigm appeared in the very beginning of the computing era, when simplicity was compulsory. Modern computer technologies, in which even mobile phones outperform the former mainframes, pose new challenges to the modeling science. A new generation of topology and state space invariant modeling paradigms is needed. They will be inevitably more structured, but not necessarily much more complex. This work is a trial step in this direction.

## Motivation

To a great extent this work was initiated by the now almost forgotten ideas of Konrad Zuse and Gordon Pask. Konrad Zuse, a German engineer who was the first to suggest that the entire universe is being computed on a computer, possibly a cellular automaton called "Rechnender Raum" or Calculating Space. In his paper he also gave an outline of a more advanced model called a net automaton (Zuse, 1969):

> *"Cellular automaton provides an elegant solution when each cell contains a complete calculating system, as symbolically represented in Figure 73. These single calculating systems contain both information-processing and information-storing elements. …The net automaton represented in Figure 74 is a further development of the cellular automaton corresponding to Figure 73. The individual cells are responsible here for only information processing. Branching lines B connect the individual cells and serve both for information transmission and for information storage."*

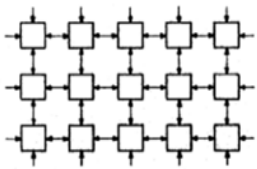 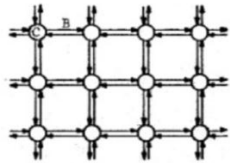

Fig. 73.   Fig. 74.

In his subsequent book on the theory of net automata (Zuse, 1975), which seems to have never been translated from German, Zuse considered mainly topological issues of the net automaton, but its internal structure was not elaborated. A decade earlier, Gordon Pask, a British cybernetician created a number of maverick machines, worked on the electro-chemical device now known as Pack's Ear (Cariani, 1993). Pask demonstrated that such a device was able to construct its own sensors and effectors without having programmed them into a preset purpose. Using the same approach, he introduced an evolutionary model containing a diffusion network, which can be regarded as a precursor of the kinetic networks considered further (Pask, 1961):

> *"All a mean by a diffusion network is a system of tubes and basins, say, over which we can define food neighborhoods. A formal representation of a food diffusion network is shown in Fig. 1. It is a directed graph with nodes. The lines connecting nodes have quantities associated with them that represent the food impedance, the amount of resistance to the passage of food between nodes."*

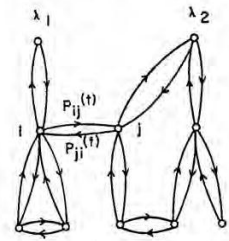

FIG. 1. Formal representation of food distribution network.

Despite the seeming disparity, both Zuse's net automaton and Pask's diffusion network have the central idea that nodes of the network are connected reciprocally with lines, having not only transport but also processing functions, and thus being active. This idea is in sharp contrast with the conventional view on network links as passive elements, prevailing until now. The main conundrum for the unification of these models was a mesh topology, because all of the earlier considered computational paradigms require a fixed number of input links, so they cannot be applied to a mesh.

The origin of this issue is that all the models, except for LGA/LBM, are *functional*, because they are based on functions with a single output value. Unlike other models, LGA/LBM models are *relational*, because they are based on relations, which are morphisms of a set of input values onto a corresponding range of output values. Relations in these models are based on equations describing the motion of fluids or gases. In order to make the models computationally tractable, these equations were calculated for a fixed number of inputs and outputs, and the results were implemented as lookup tables. Although the attempts to enhance the geometrical flexibility of LBM continue, they do not have sufficient universality (Ubertini and Succi, 2008). LGA/LBM approach seemed to be very promising, but a more universal, not restricted to a specific underlying grid or physical reality, transformation method was needed. Further investigations eventually led to a new transformation technique and a based on it model, which can be described as *a relationist view on interacting systems*.

Relationism is a venerable paradigm which can be traced back to Newton's great rival Leibniz, who argued that all properties arise from relations and reality consists of an evolving network of relationships. Later this idea was formulated by Einstein as Mach's principle. In modern physics this approach is revived by Lee Smolin, one of the leading proponents of loop quantum gravity (Smolin, 1997). This approach is also in line with the views of Gregory Bateson, a British anthropologist and cybernetician, who emphasized that logic and quantity are inappropriate devices for describing organisms and their interactions (Bateson, 1969):

> *"It is impossible, in principle, to explain any pattern by invoking a single quantity. But note that a ratio between two quantities is already the beginning of pattern. In other words, quantity and pattern are of different logical type and do not readily fit together in the same thinking. (p.53) ... We use the same words to talk about logical sequences and about sequences of cause and effect... But the if-then of logic in the syllogism is very different from the if-then of cause and effect... The if-then of causality contains time, but the if-then of logic is timeless. It follows that logic is incomplete model of causality (pp.56-59)".*

The difference between a system of causal entailments (what is happening in the external world) and a system of inferential ones (a language in which these events are expressed) was also underlined by the founder of relational biology Robert Rosen (Rosen, 1991). He defined the relation between them via further semantic elements: encoding and decoding, that bring the two entailment structures into congruence (Fig.1c).

Rosen's modeling relation is very close to a cybernetic perception-action loop (Fig.1d), which has become a conceptual schema of a general system, also known as input-process-output plus storage (IPO+S) model. From this point of view, all considered models can be divided into two main groups: *functional models* (CA/RBN/CML) (Fig.1a) and *relational models* (LGA/LBM) (Fig.1b). The main difference between them is that in functional models a new state of a cell is stored and relayed (fanned out) to all or some of its neighbors, so they can be conservative only in special cases (Fredkin and Toffoli, 1982). In contrast, relational models were created for modeling real physical phenomena, so they are conservative by default. Relational models treat a cell as a "black box" and its responses are only observable to its neighbors but not its internal state. So it can be viewed as a *reflexive system* differentially responding to its links. Its response (observable reflex) to a link depends on but is not equal to its current state and inputs from other links.

It should be noted that the range of functional models is not restricted to the listed above. All kinds of artificial neural networks, e.g. Spiking Neural Networks (Maass, 1997), and many other dynamical networks with continuous states, e.g. Compositional Pattern Producing Networks (Stanley, 2007), are also functional, because their nodes produce a single output value which is relayed only to some of their neighbors, thus they are not conservative and reflexive.

Rosen's modeling relation, cybernetic and lattice gas models were taken as a blueprint for the kinetic automaton model (Fig.1e), which inherits many of their structural and semantic elements.

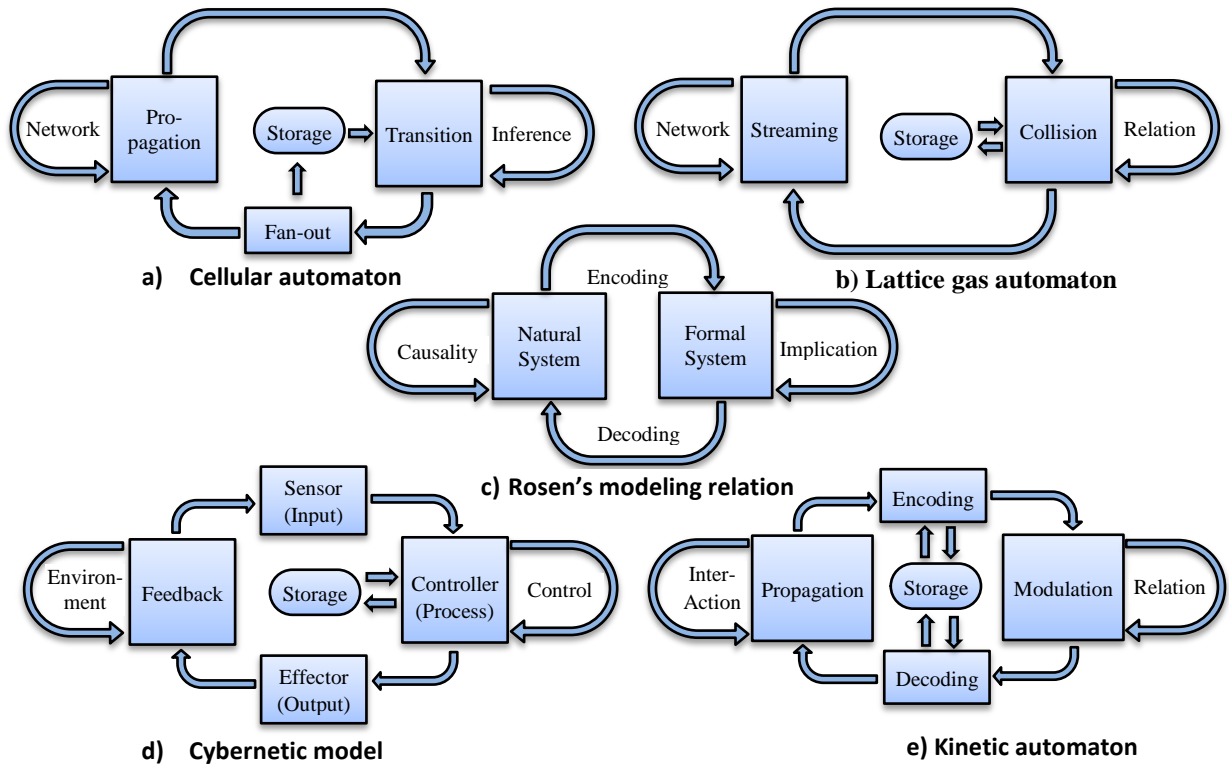

*Figure1: Modeling frameworks*

# Method

The key element of the model, making its properties and dynamics different from LBM, is a collision step, which is transformed into a 3-step operator: Encoding-Modulation-Decoding, called *Conservative Rank Transform* (CRT). In this method, not quantities as such but their relative values (ranks) are transformed (modulated), and the total quantity does not change after transformation.

CRT is derived from the simplest of all image processing techniques called Intensity Transformations (Gonzalez and Woods, 2008). The main goal of intensity transformation is to increase intensities of some pixels in an image and/or to decrease intensities of others. This is done via an intensity transformation function, which is a mapping of an input value of a pixel onto its output value (Fig.2).

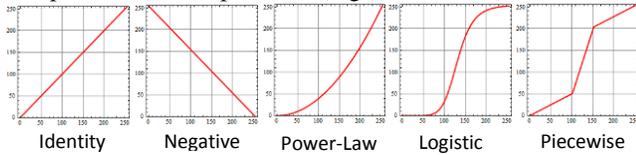

*Figure 2: Basic intensity transformation functions*

There is an advanced image processing technique, called Rank Transformation (Zabih and Woodfill, 1993), which changes the intensity of a pixel in relation to its neighborhood. CRT can be outlined as a quantity conserving synthesis of Intensity and Rank Transformations and is carried out in 3 steps:

**1. Encoding.** The first part of this step, called gathering, is just adding the values of all inputs and storage. The second one, scaling, is the conversion of absolute input values into ratios with a unit sum. In the simplest case it can be ordinary weighting, but in general, it may be a more sophisticated procedure. It should be emphasized that output values must be in the [0,1] range with a unit sum, so they can be regarded as density ratios, probabilities or ranks.

**2. Modulation.** This is the core step of the method congruent to a collision step in LBM. It modulates (maps) input ratios onto their output values via a kinetic function (map), which is a *normalized* version of the intensity transformation functions in image processing. Modulation of input ratios is performed on a one-by-one basis, so in a general case, the sum of modulated ratios changes after this step. But this does not contradict to the name of the method. Modulation only changes the proportions between input values. A kinetic map can be any smooth or piecewise curve whose domain and range are defined in the [0,1] interval.

**3. Decoding.** This step consists of the rescaling of modulated ratios to the volume of storage and scattering them among corresponding outputs and storage.

To illustrate how the method works, a simple numerical example is considered in Figure 3. During an encoding step, a set of input values {0.2, 0.4, 1.4}, denoted by *I*, is weighted to the sum of the set. The weights can be regarded as ranks with a unit sum and are denoted by *R*. On a modulation step, the ranks are modulated via a parabolic kinetic map. The sum of the modulated ranks, denoted by *R'*, increases from 1.0 to 1.6. During a decoding step for obtaining output values, denoted by *O*, each modulated rank is multiplied by a rescaling factor which is the ratio of the sum of inputs to the sum of modulated ranks (2.0/1.6=1.72). Resultant outputs have the same sum as inputs but different values.

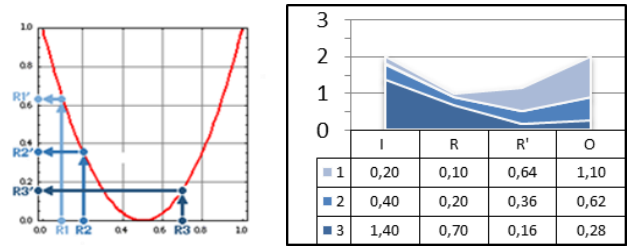

*Figure 3: Conservative Rank Transform (CRT)*

In a nutshell, this method is a quantity conserving relation or a morphism of a set of input values onto a range of corresponding output values.

# Model

The overall schema of the kinetic automaton in detail looks as follows (Fig.4):

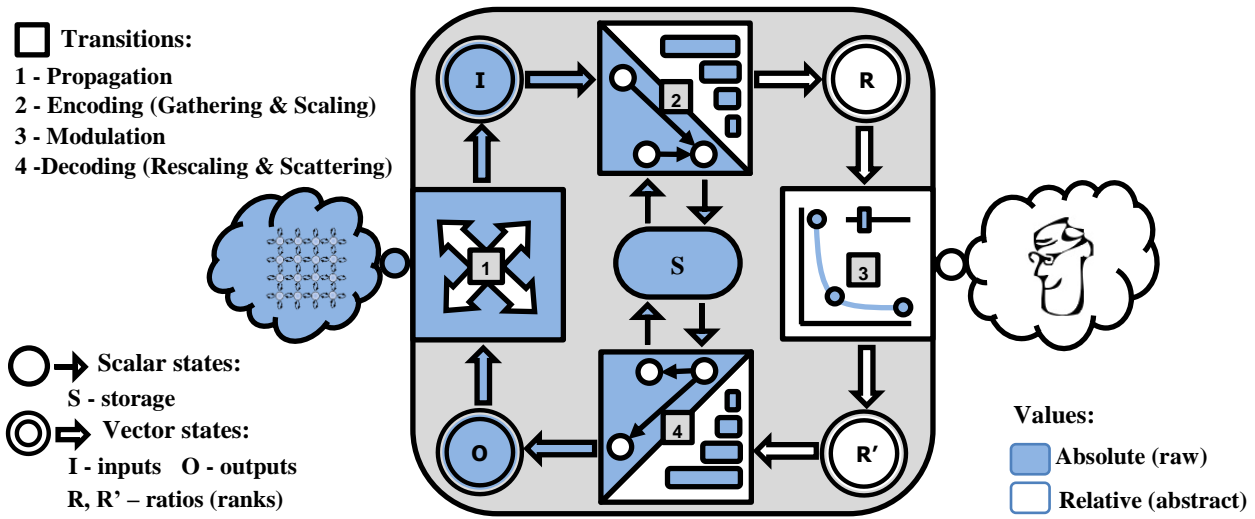

*Figure 4: The kinetic automaton state-transition structure*

This diagram uses the graphical notation of state-transition structures introduced by Carl Adam Petri and known as Petri nets (Petri, 1982). States in the kinetic automaton model can be thought of as internal buffers for storing intermediate results and are depicted by round shapes, while transitions have square forms. Input ($I$) and output ($O$) state buffers of a kinon are represented by real valued vectors of dimension $N$, where $N$ is the number of neighbors which is fixed for structured and variable for unstructured networks. Rank buffers ($R$ and $R'$) have an additional dimension corresponding to storage. The only buffer in a kinon having a scalar value is storage. It stores the overall quantity during modulation and a quantity remaining in a kinon (its state) during propagation. The sum of output storage values of all kinons of the network can be related to the internal (potential) energy of the network and the sum of output buffers of all kinons to its external (kinetic) energy. The total energy in the network is always the same, whereas internal and external energies interchangeably fluctuate.

In the diagram, a propagation block, operating with absolute values, and storage, holding an absolute (raw) value, are painted in blue (fluid) color. A modulation block operates with relative (abstract) values, so it is painted in white. Encoding and decoding blocks are proxies which manipulate with both types of values, so they are divided into two sub-blocks and painted in different colors.

In Rosen's modeling relation (Rosen, 1991) (Fig.1c), the relation of equivalence is the following:
$$1 = 2 + 3 + 4,$$
where: 1-Natural system, 2-Encoding, 3-Formal system, 4-Decoding.

If we denote divided blocks in Fig.4 as 2N, 2F, 4N, 4F and storage as S, then the following relation of equivalence between the absolute (Natural) and the relational (Formal) parts of the model holds:
$$1 + 2N + S + 4N = 2F + 3 + 4F.$$
This relation corresponds to Zuse's net automaton, where the left part designates branching lines responsible both for transmission and storage, and the right part is related to the nodes responsible for only information processing. It indicates *the dualism* of the model, because it represents both a cellular and a net automaton in Zuse's terms.

It is easy to notice the similarity of the kinetic cycle to the Carnot cycle and the engines performing it (Fig. 5).

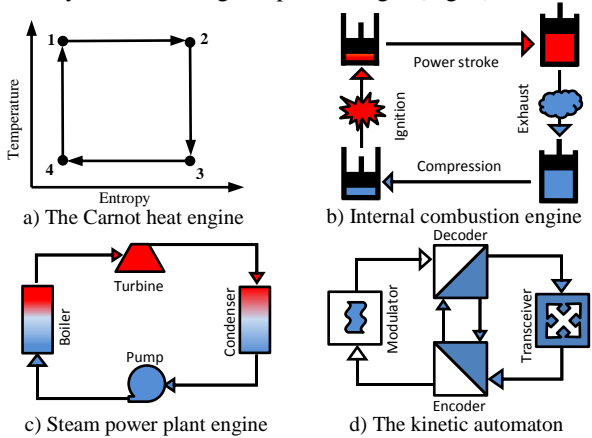

a) The Carnot heat engine  b) Internal combustion engine
c) Steam power plant engine  d) The kinetic automaton

*Figure 5: The Carnot cycle engines*

This is not a coincidence, because the model was elaborated with having the Carnot cycle as well as Kauffman's work-constraint cycle (Kauffman, 2000) in mind. The work-constraint cycle links the ideas of 'work' and 'constraint', defining work as *"the constrained release of energy into relatively few degrees of freedom"*. A system performs the work-constraint cycle if it is able to use its work to regenerate at least some of the constraints that make work possible. In Kauffman's words, *"Work begets constraints beget work"*. The kinetic automaton was conceived as an abstract autonomous agent doing its own thermodynamical cycle. Work in it is related to the kinetic exchange among automata (interaction) that generates new constraints (relation) inside automata, which in their turn, generate new interaction. Hence, the work-constraint cycle can be put as the interaction-relation cycle: *interaction begets relation begets interaction*.

The Carnot heat-engine cycle is totally reversible. Therefore, all the processes that comprise it can be reversed, in which case it becomes the Carnot refrigeration cycle. The similarity between the kinetic automaton and the Carnot cycle may lead to the conclusion that the kinetic automaton is also reversible, but it is not true. As it was mentioned earlier, the staple feature of the CRT method is its invariance to the number of inputs and outputs achieved due to decoding. Before scattering, we need to rescale the sum of modulated ratios to the volume of storage, which is an irreversible operation. In Boolean algebra, where states are binary and the number of different combinations is finite, reversibility is possible in special cases (Toffoli, 1980). In real valued algebra it is impossible in principle to split the sum without prior knowledge of components or their ratios, because the number of possible combinations giving the same result is infinite even for two components.

Therefore, this method is in line with both the conservation law and the second law of thermodynamics, which is held to be accountable for the irreversibility of time. It sounds discouraging for the model supposed to generate dynamical patterns. However, despite the second law, Life itself, which is the most improbable and fascinating of all dynamical patterns in the Universe, exists due to the phenomenon of self-organization or "*order for free*" (Kauffman, 1995). It will be shown further that in most cases the dynamics of kinetic automata networks converges to the total equilibrium or exhibits chaotic behaviour, but in some cases, even subtle changes of the kinetic map can lead to the appearance of stable or periodic patterns from an almost homogeneous state.

The kinetic automaton model can be regarded as a generalized Lattice Boltzmann model, which is not restricted to the Boltzmann equation and a regular grid, but the differences are more than that. Contrary to all considered above models, not quantities as such but their relative values are transformed, which makes the model *relational* in both Rosen's *semantic* and Mach's *relativistic* sense.

## Results

The topological universality of the model allows any network structure, but for the ease of the visualization of generated patterns, the Cartesian grid was chosen as the underlying network. It makes the conversion of the system's current state to an image quite straightforward and computationally efficient.

The elementary one-dimensional case will be considered first for better understanding of the basic principles governing the dynamics of the model. A one-dimensional kinetic network consists of *N* kinons connected in a ring. In all experiments shown below, the total quantity available in the network is equal to *N/2*. This is equivalent to the average value 0.5 in a uniform state, which is visualized as a grey color in a greyscale image. Corresponding kinetic maps are shown in the left of the images. All shown kinetic maps are defined by splines with a variable number of control points and interpolation order and both axes have [0,1] range.

It is obvious that an identity map always produces a dynamical steady state where nothing changes. However, it seems counterintuitive that an identity map is only one of many others with the same behaviour (Fig. 6a).

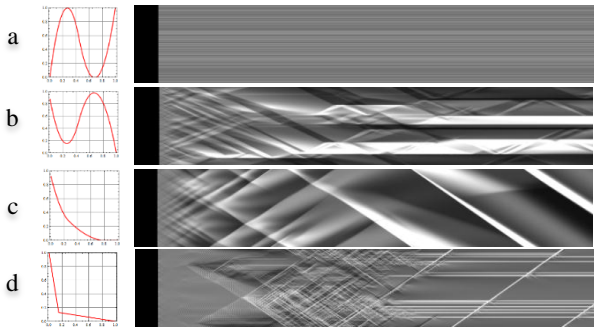

*Figure 6: 1D patterns triggered by near equilibrium states*

Generally, the dynamics of the model slowly converges to the total equilibrium or produces chaotic behaviour. However, some kinetic maps, in accordance with Turing's idea, do generate a non-uniform steady state (Fig. 6b) or give rise to travelling waves (Fig. 6c) behaving like solitons and only slightly changing after collisions. Even more, generated standing and travelling waves can coexist (Fig. 6d).

Now we change initial settings to the opposite extreme, where only one kinon has a storage which is equal to the total quantity of the network, so it can be termed as a singularity. Remarkably, the dynamics remains almost the same (Fig.7).

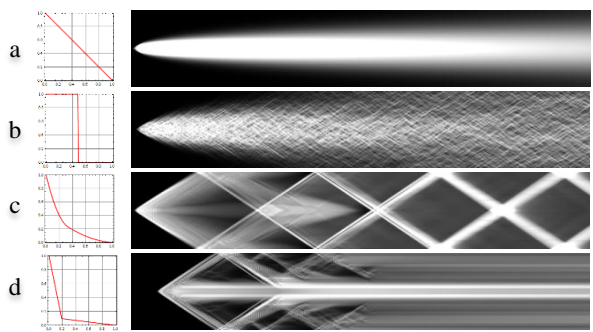

*Figure 7: 1D patterns triggered by singularity*

It is not surprising that configurations, where the storage of all kinons in the network is randomly set in the [0,1] range, behave similarly. Except for a totally uniform state, which cannot be changed by any kinetic map, the dynamics can be very sensitive to the tuning of the kinetic map in some cases and almost indifferent in others. It means that under certain conditions there exists a phase transition, which typically refers to a very narrow transition domain from one macroscopic state of the system to another. It is characterized by a sudden change in some order parameter $\varphi(\mu)$ that depends on some control parameter $\mu$ (e.g., temperature, density, probability, etc.) that can be continuously varied (Solé, 2011). Although relevant order and control parameters of the model are yet to be identified and studied, but experiments with a very simple kinetic map shown in Figure 8 revealed the existence of a narrow range of parameter *k* when a nearly uniform initial configuration converges to a dynamical ordered or chaotic pattern (*II: 1.5<k<2*) rather than a uniform (*I: k<1.5*) or non-uniform stable (*III: k>2*) state. Picture in the bottom of Figure 8 shows the change of dynamics under periodic increases of parameter *k*. The most dramatic change occurs at *k=1.6* when a nearly uniform state triggers spontainiosly into an ordered periodic pattern.

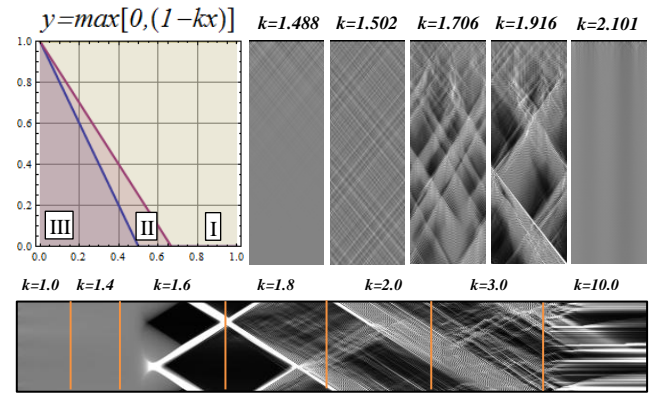

*Figure 8: One-dimensional phase transitions*

The long history of cellular automata research demonstrates that it is very difficult to produce round shapes on a square grid, but in kinetic automata it goes without any effort. In the simplest 2D configuration with a singularity and a negative kinetic map after several initial steps, an expanding wave quickly changes its form from a diamond to an ideal circle. The boundary slowly dissolves into a halo and the system converges to the total equilibrium in the long run (Fig. 9)

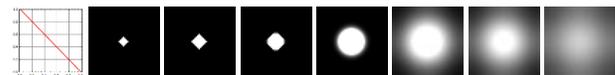

*Figure 9: Kinetic "Big Bang"*

It was shown in Figure 8 that some changes of a kinetic map can induce a sharp phase transition. If we rerun the previous experiment and, after about 150 steps, when a full-fledged circle is formed, change the kinetic map parameter *k=1.0* for *k~4.0*, an expanding circular wave splits into four sectors which begin travelling in opposite diagonal directions (Fig. 10). In all shown experiments, a kinetic map is the same for all kinons in the network and is changed manually, but there are many ways to subject kinetic map parameters to the current state of a kinon and/or its inputs, which can make its dynamics even more complicated and unpredictable.

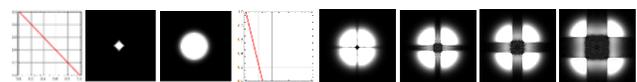

*Figure 10: Kinetic fission*

If we choose a parabolic kinetic map shown in Figure 3 and start from a singularity again, then fascinating kaleidoscopic ornaments will change non-repetitively for a long time. The picture will be even more fascinating with four evenly spaced singularities (Fig. 11).

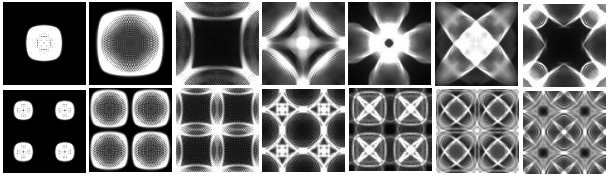

*Figure 11: Kinetic kaleidoscope*

Some kinetic maps can produce dynamic patterns which converge to a stable non-uniform state in conformity with the Turing two-factor sytems, but unlike Turing patterns, these states can be very intricate (Fig. 12):

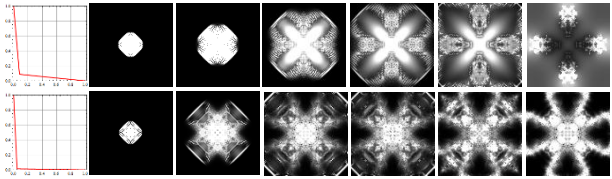

*Figure 12: Kinetic "Still Life"*

In most case studies starting from a nearly homogeneous state, a uniform grey square is usually observed, but now and then the uniformity breaks out in spots, which then turn to the chaotic or non-uniform stable state (Fig. 13)

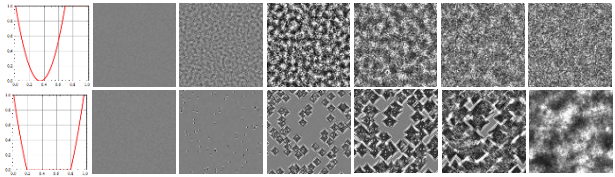

*Figure13: Chaotic and stable kinetic patterns*

It seems that it is more than enough for such a relatively simple model and it is highly unlikely to expect more, but in fact, its creativity is endless. In rare cases, one can observe almost improbable life-like phenomena, which can move, grow, merge, split and dissolve like Conway's Game of Life dynamical patterns: gliders, puffer trains, avalanches, etc. (Fig. 14)

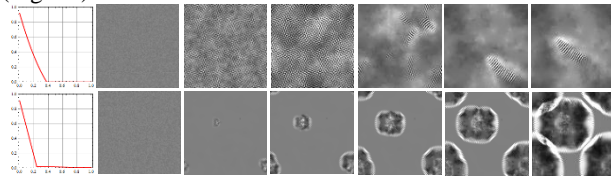

*Figure 14: Kinetic Game of Life*

## Discussion

These examples, which are an infinitely small portion of all possible patterns, prove the main thesis of this paper that even one-component kinetic networks can generate stationary and dynamic patterns similar to multi-component Turing systems. Unlike Turing systems, they are triggered by a single instability, tentatively called *kinetic instability*, which is an instance of a general *self-organized instability* (Solé et al. 2002). Multi-component kinetic networks are also possible and can be implemented by layering of all values in kinon buffers into vectors of any dimension. Apart from the same or different kinetic maps governing the dynamics of each layer there can be additional coupling among layers ('kinetic chemistry') which can be also implemented via a kinetic map. The state and dynamics of three-component kinetic networks can be readily visualized by color images.

Although the CRT method, described above and used as a computational kernel in all shown experiments, proved to be elegant and expressive, it is only one of many other possible transformation methods. The main requirement of the model is *quantity conservation* during the mapping of multiple input values and storage onto their output (feedback) values.

Closer examination revealed many unusual phenomena not found in Turing systems. In all case studies starting from a singularity, the initial diamond expanding wave, determined by the underlying square grid, quickly converges to a circle. One of the possible explanations can be done with the use of the concept of Brillouin zones. Léon Brillouin, a French physicist who studied wave propagation in periodic structures (Brillouin, 1946), introduced the concept of zones named after him. Physically, Brillouin zone boundaries represent Bragg planes which reflect waves having particular wave vectors, so that they cause constructive interference. The constriction of the first two Brillouin zones on a square grid and the final $10^{th}$ Brillouin zone is shown in Figure 15.

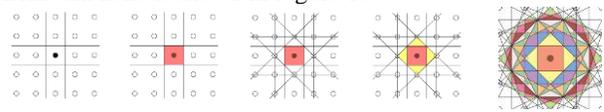

*Figure 15: Brillouin zones*

The most enigmatic phenomenon observed is an almost instantaneous fission of the expanded wave into four sectors after the change of the kinetic map (Fig.10). There is no acceptable explanation for it so far, but without a doubt, it is a self-organized phenomenon, because it requires a large-scale correlation.

Some generated patterns (Fig.11 and 12) demonstrate striking similarities with real-world phenomena, e.g., periodic patterns obtained by a Swiss medical doctor and natural scientist Hans Jenny, who coined the term *Cymatics*, the study of wave phenomena and vibrations (Jenny, 2001). Jenny pioneered the use of piezoelectric crystals hooked up to amplifiers and frequency generators to make the resultant nodal fields visible by spreading a fine powder of lycopodium spores, as well as many other materials (Fig. 16). The shapes, figures and patterns appeared to be primarily a function of frequency, amplitude, inherent properties of the materials and the size of the plate.

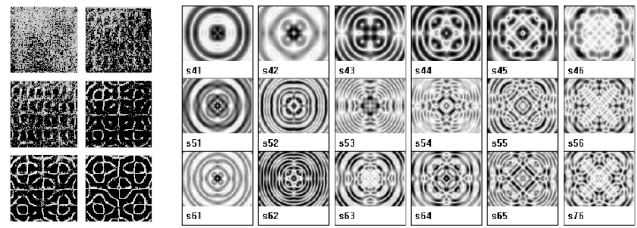

*Figure 16: Cymatic patterns*

## Conclusion

The presented results convincingly show the fidelity of the model and demonstrate high potential for future research. According to Stephen Wolfram, *"In continuous cellular automata it takes only extremely simple rules to generate behaviour of considerable complexity"* (Wolfram, 2002). It was shown here that relational approach dramatically increases the complexity of the behaviour of the kinetic automaton model. Remarkably, the proposed model structurally coincides with the recently suggested Bayesian model of perception (Friston et al., 2012), although these models are based on different assumptions and approaches. This confirms the universality and wide applicability of the model. It was demonstrated that kinetic automata possess innate tunability, which makes them a candidate toy model and a playground for studies in self-organization in general and guided self-organization (Prokopenko, 2014) in particular. The analog nature of the model enables its direct implementation with analog circuits, so the advent of memristive technologies (Adamatzky and Chua, 2014) and the revival of analog computing (Mills, 2008) may lead the way for kinetic automata to tunable kinetic media.